\documentclass[aps,twocolumn,showpacs,preprintnumbers,amsmath,amssymb]{revtex4}

\usepackage{graphicx}
\usepackage{dcolumn}
\usepackage{bm}
\usepackage{natbib}
\usepackage{braket}
\usepackage{units}

\usepackage{ae}
\usepackage{amssymb}
\usepackage{amsmath}
\usepackage{dsfont}

\newcommand{\e}{\text{e}}
\renewcommand{\d}{\text{d}}
\newcommand{\pr}{{0\nu\beta\beta}}
\newcommand{\bma}{\begin{pmatrix}}
\newcommand{\ema}{\end{pmatrix}}
\newcommand{\be}{\begin{equation}}
\newcommand{\ee}{\end{equation}}

\begin{document}

\preprint{DO-TH-10/14}
\preprint{TUM-HEP-798/11}

\title{Fourth Generation Majorana Neutrinos}

\author{Alexander Lenz}
 \email{alexander.lenz@cern.ch}
\affiliation{Physik Department, Technische Universit\"at M\"unchen, D-85748 Garching, Germany}
\author{Heinrich P\"as}
 \email{heinrich.paes@uni-dortmund.de}
\author{Dario Schalla}
 \email{dario.schalla@tu-dortmund.de}
\affiliation{Department of Physics, Technische Universt\"at Dortmund, D-44221 Dortmund, Germany}

\date{May 2, 2012}

\begin{abstract}
We investigate the possibility of a fourth sequential generation in the lepton sector. Assuming neutrinos to be Majorana particles and starting from a recent -- albeit weak -- hint for a nonzero admixture of a fourth generation neutrino from fits to weak lepton and meson decays, we discuss constraints from neutrinoless double beta decay, radiative lepton decay, and like-sign dilepton production at hadron colliders. Also, an idea for fourth generation neutrino mass model building is briefly outlined. Here we soften the large hierarchy of the neutrino masses within an extradimensional model that locates each generation on different lepton number violating branes without large hierarchies.
\end{abstract}

\maketitle


\section{Introduction}
Adding another family of fermions to the three known generations
has become a popular extension of the standard model (SM3); see e.g.
\cite{Frampton:1999xi, Holdom:2009rf} for reviews. It is now common 
knowledge that a fourth family is not excluded by electroweak precision constraints
\cite{Holdom:1996bn,Maltoni:1999ta,He:2001tp,Novikov:2001md,Kribs:2007nz,Novikov:2009kc,Erler:2010sk,Eberhardt:2010bm,Chanowitz:2010bm}
or by flavor constraints
\cite{Bobrowski:2009ng,Soni:2010xh,Buras:2010pi,Hou:2010mm,Nandi:2010zx,Alok:2010zj}.
Moreover, this extension (SM4) of the standard model also offers several desired features:
\begin{itemize}
\item The SM4 can weaken the tension between direct and indirect bounds on the Higgs mass
      \cite{Novikov:2002tk,Frere:2004rh,Kribs:2007nz,Novikov:2009kc,Hashimoto:2010at}.
\item The SM4 can lead to a sizable enhancement of the measure of 
      CP violation \cite{Hou:2008xd,Hou:2010wf} and therefore help 
      to solve the problem of baryogenesis. In addition, the
      strength of the phase transition might also be increased    
      \cite{Carena:2004ha,Fok:2008yg,Kikukawa:2009mu}.      
\item If problems with arising Landau poles can be cured, the SM4 might also help to achieve a unification of couplings \cite{Hung:1997zj}.
\item Large Yukawa couplings might lead to new, interesting, strong dynamic effects
      \cite{Hung:2009hy,Hung:2009ia} including also dynamical symmetry breaking
      \cite{Holdom:1986rn,Carpenter:1989ij,King:1990he,Hill:1990ge,   
      Hung:1996gj,Holdom:2006mr,Burdman:2008qh,Hashimoto:2009ty,Antipin:2010it,Delepine:2010vw,Simonov:2010wd,Fukano:2011fp,Hung:2010xh}.
\item Despite the enormous success of the Cabibbo-Kobayashi-Maskawa picture \cite{Cabibbo:1963yz,Kobayashi:1973fv}, 
      the assumption that both $B_d$ and $B_s$ mixing are described by the SM3 
      alone is excluded by 3.8 standard deviations \cite{Lenz:2010gu}.
      Some of these problems in flavor physics might be cured by the SM4; see
      \cite{Soni:2009fg,Soni:2008bc,Hou:2006mx,Arhrib:2006pm,Hou:2005hd,Choudhury:2010ya} for some recent work and     
      e.g. \cite{Hou:1986ug,Hou:1987vd} for some early work on fourth generation effects on     
      flavor physics. This topic gained a lot of interest recently due to
      the measurement of the dimuon asymmetry from the D0 Collaboration \cite{Abazov:2010hv,Abazov:2010hj}
      which was a factor of 42 larger than the SM3 result \cite{Lenz:2006hd,Nierste:2011ti} - with a statistical significance
      of 3.2 standard deviations. The SM4 can also enhance the dimuon asymmetry considerably,
      albeit not by a factor of 42.
\end{itemize}
Besides all these promising facts there is one property of the SM4 which is typically considered 
to be very : the masses of the neutrinos of the first three families are 
below ${\cal O} (1 \, \mbox{eV})$, while the mass of the fourth neutrino has to be above 
${\cal O} (100 \, \mbox{GeV})$. In this work we try to shed some light
on this gap of at least 11 orders of magnitude.

The particle content of the SM4 model is as follows:
\begin{displaymath}
\begin{array}{lllllll}
\mbox{1st family}: & \left( u \atop d  \right)_L, & u_R, & d_R, &  \left( \nu_e \atop e^-  \right)_L, & e^-_R, & \nu_{e,R}
\\
\\
\mbox{2nd family}: & \left( c \atop s  \right)_L, & c_R, & s_R, &  \left( \nu_\mu \atop \mu^-  \right)_L, & \mu^-_R, & \nu_{\mu,R}
\\
\\
\mbox{3rd family}: & \left( t \atop b  \right)_L, & t_R, & b_R, &  \left( \nu_\tau \atop \tau^-  \right)_L, & \tau^-_R, & \nu_{\tau,R}
\\
\\
\mbox{4th family}: & \left( t' \atop b'  \right)_L, & {t'}_R, & {b'}_R, &  \left( \nu_4 \atop l_4^-  \right)_L, & l^-_{4,R}, & \nu_{4,R}
\\
\\
\end{array}
\end{displaymath}

Recent work in the leptonic sector can be found in \cite{Carpenter:2010sm,Buras:2010cp,Burdman:2009ih,Blennow:2010th,Antusch:2008tz,Antusch:2006vwa,Frandsen:2009fs,Garg:2011xg,Aparici:2011nu}.

The Dirac mass of the $i$th generation neutrino is denoted by $m_{Di}$,
while the Majorana mass is denoted by $M_{Ri}$. The light neutrino mass eigenstate is $m_i$, and the heavy mass eigenstate is $M_i$.
We  use the following experimental bounds from direct searches \cite{Amsler:2008zzb}:
\begin{eqnarray}      
m_4  & > & 80.5 ... 101.5   \, \mbox{GeV}  \, , 
\\   
m_{l_4}    & > & 100.8   \, \mbox{GeV}   \, , 
\end{eqnarray}      
These mass bounds depend on the type of neutrino (Dirac or Majorana) and whether   
one considers a coupling of the heavy neutrino to $e^-$, $\mu^-$, or $\tau^-$.   
An investigation \cite{Eberhardt:2010bm} of the oblique electroweak parameters
\cite{Peskin:1991sw,Peskin:1990zt} gives
\begin{equation}
|m_{l_4}-m_4|<140 \,  \mbox{GeV}.
\label{eq:degen}
\end{equation}
As the light neutrino mass eigenvalues are bounded from above by the Dirac masses -- at least in a typical seesaw model -- this provides a bound on the Dirac-type mass $m_{D4}$ as well. Assuming perturbativity of the fourth generation neutrino Yukawa couplings, the Dirac-type neutrino mass is approximately constrained to the interval
\be
\unit[45]{GeV} \leq m_{D4} \leq \unit[1000]{GeV}
\label{eq:interval}
\ee
where the lower bound arises from the invisible $Z$ decay width that fixes the number of neutrino generations to three for neutrino masses less than half the mass of the $Z$ boson~\cite{:2005ema}.

Finally, a recent  partial Pontecorvo-Maki-Nakagawa-Sakata (PMNS) fit to a set of experimental data in the $SM4$ framework has provided a hint for a nonzero admixture of a fourth generation neutrino ($2\sigma$ limits), resulting in a PMNS matrix~\cite{Lacker:2010zz}.
Note that these bounds have changed due to new experimental data since the original publication~\cite{Lacker:2010zz}. In particular, the matrix element $U_{e4}$ is now compatible with zero at the 2$\sigma$ level~\cite{MenzelTalk}. The central value is at $U_{e4}=0.044$ with $0.015 <U_{e4}< 0.060$ at the 1$\sigma$ level. This article studies implications of a nonvanishing matrix element in terms of order of magnitude estimations. As these new bounds are not published, we use the results of~\cite{Lacker:2010zz} as given in Eq.~(\ref{eq:LackerPMNS}). The assumptions remain well motivated, and the resulting estimations and analyses do not vary significantly with the precise values of $U_{e4}$.
\begin{equation}
U = \begin{pmatrix}
*      & *      & *      & _{>0.021}^{<0.089} \\
*      & *      & *      & <0.029 \\
*      & *      & *      & <0.085 \\
<0.115 & <0.115 & <0.115 & _{>0.9934}^{<0.9998}
\end{pmatrix} .\label{eq:LackerPMNS}
\end{equation}
In principle, a fourth generation neutrino can induce radiative contributions to the light neutrino mass eigenstates which may
exceed the neutrino mass bounds obtained from cosmology, Tritium beta decay and neutrinoless double beta decay.
At present, the cosmological bound on the two-loop contributions gives a more stringent bound than the
bound from neutrinoless double beta decay discussed here. However, as cosmological bounds suffer
from large systematic uncertainties, we restrict ourselves here to the discussion of neutrinoless double beta
decay. For a discussion of fourth generation induced loop effects on light neutrinos, we refer to~\cite{Petcov:1984nz,Babu:1988ig} and the recent papers~\cite{Aparici:2011nu,Schmidt:2011jp}. However, the findings of this article in the context of neutrinoless double beta decay remain valid on their own.

\section{Neutrinoless double beta decay}
\begin{figure}
\begin{minipage}{.45\textwidth}
\centering
\includegraphics{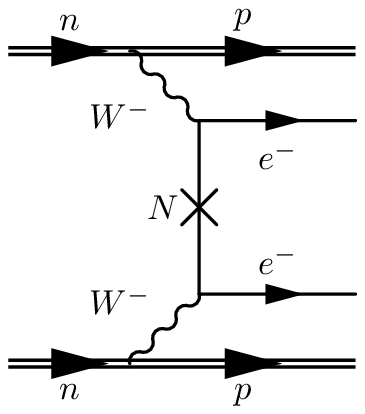}
\caption{Feynman diagram of $\pr$ induced by the exchange of a heavy fourth generation Majorana neutrino.}
\label{fig:0vbbFEY}
\end{minipage}
\begin{minipage}{.45\textwidth}
\centering
\includegraphics{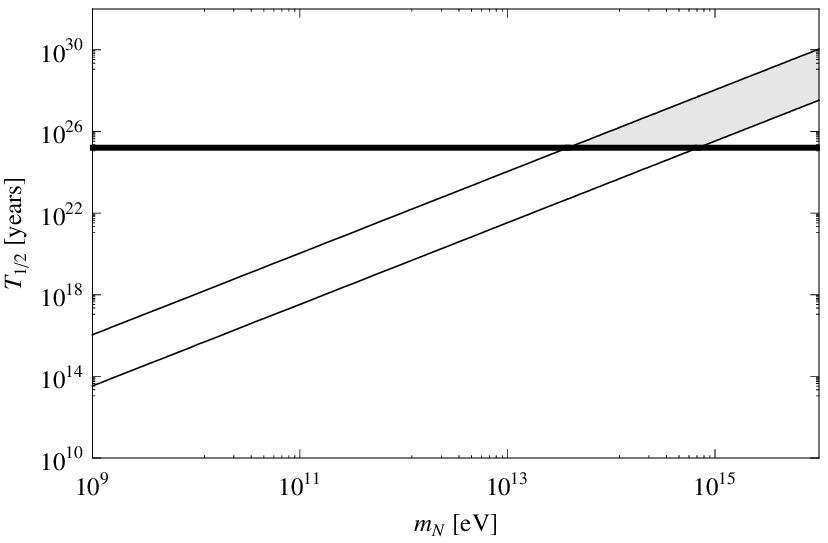}
\caption{Contribution of heavy neutrinos on a $\pr$ half-life (thin lines) within the mixing region given by Eq.~(\ref{eq:LackerPMNS}) and IGEX lower bound (thick line). The gray area indicates the allowed region.}
 \label{fig:TofM}
\end{minipage}
\end{figure}

The most sensitive probe for neutrino Majorana masses is generally
neutrinoless double beta decay ($\pr$).
$\pr$ decay can be realized by the exchange of a Majorana neutrino (see Fig.~\ref{fig:0vbbFEY}). In the presence of additional heavy neutrino states the usual effective Majorana mass $\braket{m_\nu}$ has to be complemented by an effective heavy neutrino mass $\Braket{m_N}^{-1}$:
\be
\braket{m_\nu} = \sum_\alpha U_{e\alpha}^2 m^\nu_\alpha \hspace{1cm} \Braket{m_N}^{-1} = \sum_\beta U_{e\beta}^2 (m^N_\beta)^{-1},
\ee
where $m^\nu_\alpha$ ($m^N_\beta$) are neutrino mass eigenstates lighter (heavier) than $\mathcal{O}(\unit[100]{MeV})$.
The half-life of the decay is then given by \cite{Hirsch:1996qw}
\begin{multline}
\left[ T_{1/2}^{0\nu\beta\beta} \right]^{-1} =  \left( \frac{\Braket{m_\nu}}{m_e} \right)^2 C_{mm}^{LL} + \left( \frac{m_p}{\Braket{m_N}} \right)^2 C_{mm}^{NN} \\
+ \left( \frac{\Braket{m_\nu}}{m_e} \right) \left( \frac{m_p}{\Braket{m_N}} \right)  C_{mm}^{NL},
\end{multline}
where the $C_{mm}$ factors include phase space factors and nuclear matrix elements \cite{Hirsch:1995rf,Muto:1989hw} and $m_p$ ($m_e$) is the proton (electron) mass.
Considering only the heavy neutrino contribution and using the PMNS matrix given in Eq.~(\ref{eq:LackerPMNS}), one obtains stringent bounds on the allowed mass range from the current experimental lower half-life bound $T_{1/2}^\textmd{Ge} > \unit[1.57 \cdot 10^{25}]{years}$ \cite{Aalseth:2002rf}. The allowed region is shown in Fig.~\ref{fig:TofM}.\\
This leads to the following mass bounds for a single fourth generation Majorana neutrino
\begin{eqnarray}
\label{eq:Mpure1} 
U_{e4}^{max} = 0.089 \; & \Rightarrow & \; m_4 \geq \unit[6.8 \cdot 10^5]{GeV} \label{eq:Mpure2a}\\
\label{eq:Mpure2b} U_{e4}^{min} = 0.021 \; & \Rightarrow & \; m_4 \geq \unit[3.8 \cdot 10^4]{GeV},
\end{eqnarray}
which are far above the perturbativity constraint of
 Eq.~(\ref{eq:interval}).
 
Relative phases between light and heavy contributions, introduced by
\begin{eqnarray}
\Braket{m_\nu} &\Rightarrow& \e^{i\alpha} \Braket{m_\nu}\\
\Braket{m_N}^{-1} &\Rightarrow& \e^{i\beta} \Braket{m_N}^{-1},
\end{eqnarray}
may cancel each other and thus loosen this bound. The most effective cancellation is possible if the light neutrinos are quasidegenerate with masses at the upper bound consistent with the large scale structure of the Universe~\cite{Spergel:2006hy},
\be
\sum m_\nu < \unit[0.66]{eV}.
\ee
With this assumption and using mass splittings obtained from neutrino oscillation analyses \cite{GonzalezGarcia:2010er},
the mass region of the heavy neutrino can be lowered compared to Eqs.~(\ref{eq:Mpure2a}) and (\ref{eq:Mpure2b}),
\be
m_4 \geq \unit[2.50 \cdot 10^4]{GeV} \; (\unit[4.49 \cdot 10^5]{GeV})
\ee
for $U_{e4}^{min} \; (U_{e4}^{max})$, respectively, which remains several orders of magnitude above the desired range (\ref{eq:interval}).

In principle there are three different ways to save the possibility of a heavy fourth generation neutrino:
\begin{enumerate}
 \item neutrinos are Dirac particles and therefore $\pr$ is forbidden,
which would come at the cost of seesaw neutrino mass suppression and
leptogenesis as a successful way to generate the baryon asymmetry of the Universe;
 \item some other physics beyond the standard model is involved and cancels the heavy neutrino contribution, which would require fine-tuning;
 \item the fourth generation neutrinos are pseudo-Dirac particles.
\end{enumerate}
In the following, we will focus on the 
latter alternative, which may provide useful guidance for future model 
building.\\
Pseudo-Dirac neutrinos arise when the Majorana mass is small compared to the Dirac mass. The two resulting mass eigenstates ($m_4$, $M_4$) are nearly degenerate 
with tiny mass splitting $\delta m$,
and the active and sterile components exhibit practically maximal mixing:
\be
\tan 2\theta \approx \frac{2 m_{D4}}{m_4 - M_4}.
\ee
Because of their opposite creation phases, the contributions of the two individual fourth generation neutrinos to $\pr$ cancel each other and only the arbitrary small mass difference contributes to the effective mass. This has already been introduced as a mechanism to hide light neutrinos in $\pr$ \cite{Beacom:2003eu}. We use this idea to hide a fourth generation Majorana neutrino with a mass of the order of the electroweak scale.

The $\pr$ half-life of a heavy pseudo-Dirac neutrino reads
\be
\left[ T_{1/2}^{0\nu\beta\beta} \right]^{-1} = \frac{1}{4} \left( \frac{m_p}{\Braket{m_4}} -  \frac{m_p}{\Braket{M_4}} \right)^2 C_{mm}^{NN} .
\label{eq:0vbbPD} 
\ee

Hence the maximal allowed mass splitting
\be
\delta m = M_4 - m_4
\ee
has to be small enough to compensate the large contribution of each individual neutrino.

\begin{figure}
 \centering
 \includegraphics{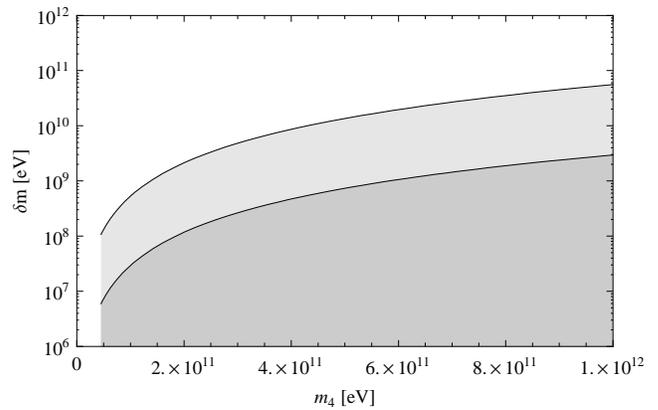}
 \caption{Maximal mass splitting for heavy pseudo-Dirac neutrinos. The upper (lower) curve corresponds to the lower (upper) bound on $U_{e4}$ according to Eq.~(\ref{eq:LackerPMNS}). The shaded area represents the allowed parameter space.}
 \label{fig:Mdiff}
\end{figure}

The required maximal mass splittings are shown in Fig.~\ref{fig:Mdiff}.

The limits for $U_{e4}^{min}$ imply $\delta m \leq \unit[107]{MeV} (\unit[56]{GeV})$ for $m_4 = \unit[45]{GeV} (\unit[1000]{GeV})$, respectively.

The ratio of the mass difference and absolute mass scale $\delta m / m_4$ is thus of the order of $10^{-3}-10^{-2}$.

\section{Radiative lepton decays}
The analysis on neutrino mixing carried out in \cite{Lacker:2010zz} is dominated by
the radiative lepton flavor violating decays of charged leptons.

\begin{figure}
\centering
\includegraphics{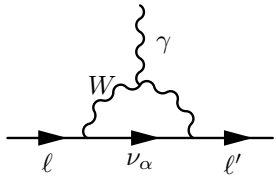} 
\caption{Feynman diagram of the radiative decay of a lepton.}
\label{fig:radiativedecay}
\end{figure}

The processes (see Fig.~\ref{fig:radiativedecay}) and their experimental bounds \cite{Adam:2011ch,Aubert:2009tk} are
\begin{eqnarray}
BR ( \mu\rightarrow e \gamma ) &<&  2.4 \cdot 10^{-12} \\
BR ( \tau\rightarrow \mu \gamma ) &<& 4.4 \cdot 10^{-8} \\
BR ( \tau\rightarrow e \gamma ) &<& 3.3 \cdot 10^{-8} .
\end{eqnarray}

In general, the amplitudes are given by \cite{Cheng:1980tp}
\begin{equation}
 T_\alpha = U_{\ell \alpha} U_{\ell' \alpha} F\left( \frac{m_\alpha^2}{m_W^2} \right)
\end{equation}
where, defining $x_\alpha \equiv \frac{m_\alpha^2}{m_W^2}$, $F(x_\alpha )$ is
\begin{multline}
 F(x_\alpha ) = 2 (x_\alpha +2) I^{(3)} (x_\alpha ) - 2 (2x_\alpha -1) I^{(2)} (x_\alpha ) \\
+ 2x_\alpha I^{(1)} (x_\alpha ) +1
\end{multline}
with
\begin{equation}
 I^{(n)}(x_\alpha ) = \int_0^1 \d z \frac{z^n}{z+(1-z)x_\alpha }.
\end{equation}
Thus the decay width can be written as
\be
 \Gamma_{\ell\rightarrow\ell'\gamma} = \frac{1}{2} \frac{G_F^2m_\ell^5}{(32\pi^2)^2} \alpha_W \sum_{\alpha} \left| U_{\ell \alpha} U_{\ell' \alpha} \right|^2 F_{eff}^2(x_\alpha )
\label{eq:radiative}
\ee
with $F_{eff} (x_\alpha )= F(x_\alpha ) - F(0)$ due to the unitarity cancellation $\sum_\alpha U_{\ell \alpha} U_{\ell' \alpha}$ of the constant term.

Here we shortly reconsider the bound for the case of pseudo-Dirac neutrinos
with masses in the 100~GeV range.

\begin{figure}
  \centering
  \begin{minipage}{.45\textwidth}
    \includegraphics[width=\linewidth]{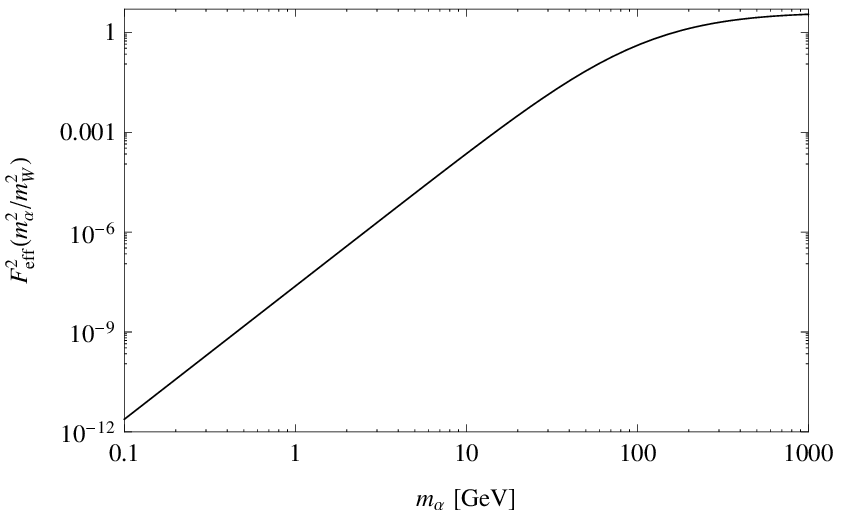} 
    \caption{$F_{eff}^2$ as a function of the mass of the exchanged neutrino.\\[1.8 cm]\hbox{}}
    \label{fig:Fsq}
  \end{minipage}
  \begin{minipage}{.45\textwidth}
  	\centering
    \includegraphics[width=.7\linewidth]{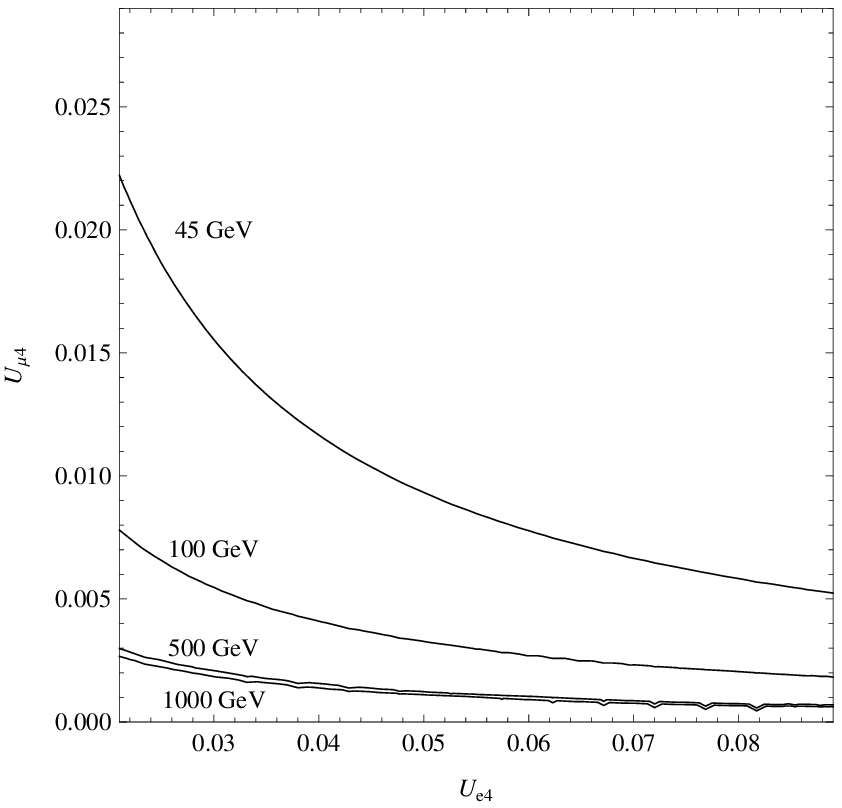}  
    \caption{Constraint on the $U_{\mu 4}-U_{e 4}$ parameter space obtained from the bound on the branching ratio for $\mu \rightarrow e \gamma$. The lower left is the allowed region. The boundaries of the intervals plotted are given by the allowed values $U_{\mu 4}-U_{e 4}$ with $m_4 = \unit[45]{GeV}$ according to Eq.~(\ref{eq:LackerPMNS}).}
    \label{fig:mu-egamma}
  \end{minipage}
\end{figure}

It is easy to see that the analysis holds for a pseudo-Dirac neutrino as well. 
As the fourth generation active and sterile states mix maximally and the masses are close to degenerate,
$F(x)$ does not change considerably compared to the pure Dirac case.
As can be seen from Fig.~\ref{fig:Fsq},
the decay rate is suppressed by the tiny masses for the first three generations, while the contribution of a fourth heavy generation has to be suppressed due to small mixing.

In the analysis of \cite{Lacker:2010zz} the neutrino  mass was fixed to 45 GeV. 
While a mass dependent study is encouraged,  the conclusions of this work
will remain unchanged, as the size of the allowed region is anticipated to
vary only slightly for different neutrino masses.
As it is obvious from (\ref{eq:radiative}), only the product $\left| U_{\ell \alpha} U_{\ell' \alpha} \right|^2$ is constrained, not one of the individual quantities alone. Since no lower bound on $U_{\mu 4}$ exists, larger masses $m_4$ corresponding to larger rates for $\mu \rightarrow e \gamma$ do not result in a more stringent bound on $U_{e4}$ (see Figs.~\ref{fig:Fsq} and \ref{fig:mu-egamma}).

The decays of the $\tau$ lepton do not provide further information
as the experimental constraints in this channel are much weaker.

\section{Like-sign dilepton production}
\begin{figure}
\centering
\begin{minipage}{.45\textwidth}
\centering
 \includegraphics{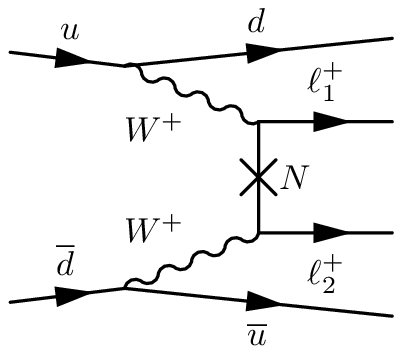}
\end{minipage}
\centering
\begin{minipage}{.45\textwidth}
\centering
 \includegraphics{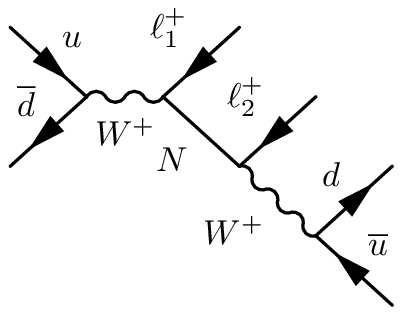}
\end{minipage}
\caption{Feynman diagrams of like-sign dilepton production}
\label{fig:likesign}
\end{figure}

\begin{figure}
 \centering
 \includegraphics{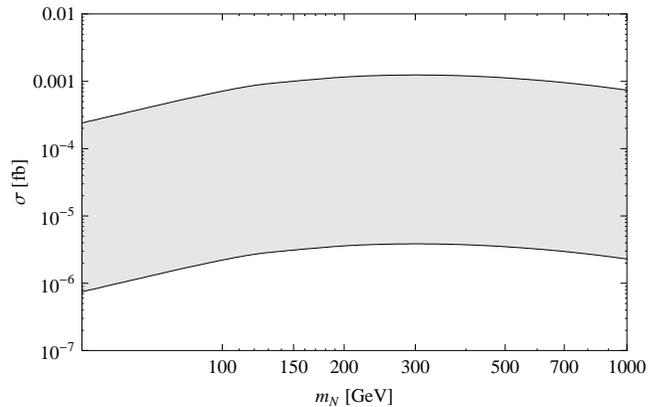}
 \caption{Cross section for like-sign dielectron production by an electroweak scale Majorana neutrino without $\pr$ constraints. The shaded area corresponds to the allowed values of $U_{e4}$ according to Eq.~(\ref{eq:LackerPMNS}).}
 \label{fig:LikeSign}
\end{figure}

\begin{figure}
 \centering
 \includegraphics{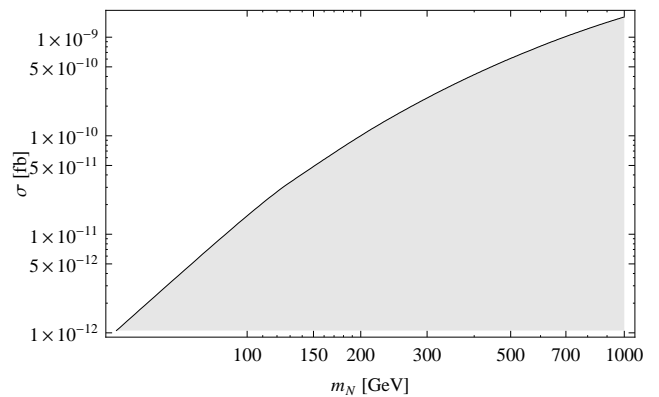}
 \caption{Upper bound on the cross section for like-sign dielectron production triggered by electroweak scale pseudo-Dirac neutrino fulfilling $\pr$ constraints.}
 \label{fig:LikeSignPS}
\end{figure}

Finally, 
a process very similar to $\pr$ is the production of two charged leptons 
of the same charge at hadron colliders:
\be
pp \rightarrow \ell_1^+ \ell_2^+ X .
\ee
As shown in Fig.~\ref{fig:likesign}, a heavy Majorana neutrino exchange drives the process whose cross section is \cite{Ali:2001gsa}:
\begin{multline}
\sigma_{single} \left( pp \rightarrow \ell^+_1 \ell_2^+ X \right) = \frac{G_F^4 m_W^6}{8\pi^5} \left( 1 - \frac{1}{2} \delta_{\ell_1 \ell_2} \right) \\
\times
\left| U_{\ell_14} U_{\ell_24} \right|^2 F \left( E, m_4 \right),
\label{eq:LSD}
\end{multline}
where $F \left( E, m_4 \right)$ is a function of beam energy and neutrino mass.
To describe the exchange of
a pseudo-Dirac neutrino, the cross section has to be modified to include the contributions of the two almost degenerate mass eigenstates in the neutrino propagator. In contrast to $\pr$ where the typical momentum of the neutrino is given by the average nuclear momentum which is smaller than $m_4$, resulting in the inverse mass dependence of Eq.~(\ref{eq:0vbbPD}), the momentum at hadron colliders exceeds $m_4$, leading to a neutrino propagator proportional to the neutrino mass. This was also shown in the relevant subprocess $W^+ W^+ \rightarrow \ell^+ \ell'^+$ \cite{Belanger:1995nh} that is included in Eq.~(\ref{eq:LSD}). To obtain the cross section $\sigma$ for a pseudo-Dirac neutrino, we now have to introduce the correction factor $\Delta_{pD}$ defined by
\be
\sigma = \Delta_{pD} \cdot \sigma_{single} \left( pp \rightarrow \ell^+_1 \ell_2^+ X \right)  .
\ee
Obviously, by comparison of the propagators, this factor is given by
\be
\Delta_{pD}  = \frac{1}{4} \frac{1}{m_4^2} (M_4 - m_4)^2 = \frac{1}{4} \left( \frac{\delta m}{m_4} \right)^2.
\ee
Here 
the mass splitting 
$\delta m$ follows from 
the $\pr$ constraint and results in a suppression of the mass dependent cross section. For example, the unsuppressed dielectron production cross section shown in Fig.~\ref{fig:LikeSign} is several orders of magnitude larger than the suppressed cross section of a pseudo-Dirac neutrino fulfilling the $\pr$ constraints shown in Fig.~\ref{fig:LikeSignPS}.

\begin{figure}
 \centering
 \includegraphics{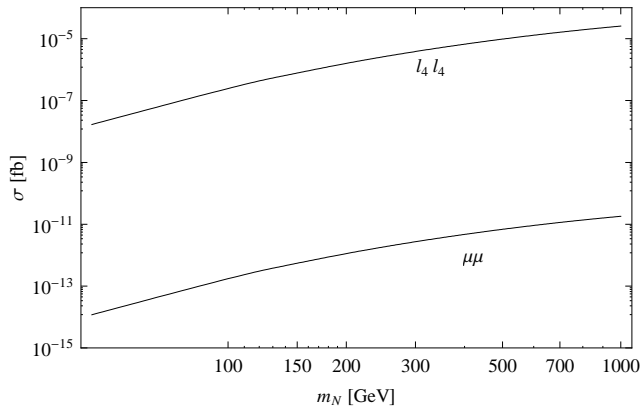}
 \caption{Cross section for like-sign dilepton production triggered by an electroweak scale pseudo-Dirac neutrino fulfilling the $\pr$ constraint (\ref{eq:LikeSignFlavor}). To obtain the cross section for specific flavors of the final state leptons, the curve has to be scaled according to the factors given in Table~\ref{tab:FlavorDependence}. The lines show the upper bound on the two extreme channels $l_4l_4$ and $\mu\mu$.}
 \label{fig:likesignFLAVOR}
\end{figure}

The cross section depends on the flavors of the $\ell_i$ leptons via the factor
\be \left( 1-\frac{1}{2} \delta_{\ell_1\ell_2}\right) \left| U_{\ell_14} U_{\ell_24} \right|^2. \label{eq:LikeSignFlavor} \ee
The unsuppressed cross section shown in Fig.~\ref{fig:likesignFLAVOR} is reduced by this factor. The upper bounds of this factor for different final flavor states calculated from Eq.~(\ref{eq:LackerPMNS}) are listed in Table~\ref{tab:FlavorDependence}. The cross section~(\ref{eq:LSD}) neglects charged lepton mass effects, but as $\sqrt{s} = \unit[7]{TeV} \gg m_{l_4}$ it is also applicable to fourth generation charged lepton production providing a reasonable upper bound of the cross section.

\begin{table}
	\centering
		\begin{tabular}{|c|c|}
			\hline
			Final state	&	Flavor factor [$10^{-5}$]	\\
			\hline
			$l_4 l_4$	&	49960		\\
			$e l_4$		&	791.78	\\
			$\tau l_4$&	722.21	\\
			$\mu l_4$	&	84.07		\\
			$e\tau$		&	5.72		\\
			$ee$			&	3.14		\\
			$\tau\tau$&	2.61		\\
			$e\mu$		&	0.67		\\
			$\mu\tau$	& 0.61		\\
			$\mu\mu$	&	0.04		\\
			\hline
		\end{tabular}
	\caption{Flavor dependent scaling factor (\ref{eq:LikeSignFlavor}) of the like-sign dilepton production cross section shown in Fig.~\ref{fig:likesignFLAVOR}. The states are sorted by their flavor factor.}
	\label{tab:FlavorDependence}
\end{table}

It is obvious that will decay predominantly into the fourth generation charged leptons. As the fourth generation neutrino mixing to the muon has the smallest upper bound, this channel has the smallest upper bound on the cross section.
However, all resulting cross sections are far too small to be observed at the expected
LHC luminosities.

\section{Mass model}
An argument often given against a fourth fermion generation is the large mass hierarchy between the first three generations and the fourth generation, in particular, in the neutrino sector. The large scale structure of the Universe implies sub-eV masses for the known light neutrinos, whereas the electroweak observables call for a fourth generation neutrino mass eigenstate of several hundreds GeV, thus creating a tension of about 12 orders of magnitude.

The simplest way to account for small neutrino masses arises from the Lagrangian,
\begin{equation}
- \mathcal{L} = \frac{1}{2} \bma \overline{\nu}_L & \overline{N}_L^C \ema \bma 0 & m_D \\ m_D^T & M_R \ema \bma \nu_R^C \\ N_R \ema,
\end{equation}
where $m_D$ is the Dirac mass matrix and $M_R$ the Majorana mass matrix.

For a single generation $i$ the resulting mass eigenstates are
\begin{eqnarray}
m_i = \frac{1}{2} \left( \sqrt{4m_{Di}^2+M_{Ri}^2} - M_{Ri} \right) \\ 
M_i = \frac{1}{2} \left( \sqrt{4m_{Di}^2+M_{Ri}^2} + M_{Ri} \right).
\end{eqnarray}
The Dirac mass is given by the Yukawa coupling $y_i$ as
$$ m_{Di} = y_i v $$
with $v\approx \unit[246]{GeV}$ the vacuum expecation value of the Higgs field.

The type-I seesaw model generates small neutrino masses by introducing large Majorana masses ($M_{Ri}\gg m_{Di}$), leading to the mass eigenstates
\begin{equation}
 m_i \approx \frac{m_{Di}^2}{M_{Ri}} \hspace{1cm} M_i \approx M_{Ri}.
\end{equation}

Then, again, for the fourth generation pseudo-Dirac neutrino the Majorana mass $M_{R4}$ is constrained to roughly the MeV scale (see Fig.~\ref{fig:Mdiff}):
\begin{equation}
 \delta m = M_4 - m_4 = M_{R4}.
\end{equation}

Thus, to satisfy all neutrino generations one has to introduce large hierarchies of either the Yukawa couplings or the Majorana masses. 
As the Majorana masses are not affected by electroweak symmetry breaking, it is more natural to assume the generation of the hierarchy in this sector rather than in the Yukawa sector.

The hierarchy reaches from MeV scale of the fourth generation up to the grand unification scale ($\sim \unit[10^{16}]{GeV}$) needed for the light neutrinos when assuming similar Yukawa couplings for all generations.

This hierarchy can be considerably softened in the context of extra spatial dimensions. It is known that approximate symmetries on our SM brane can be broken at a different brane located at some distance in the extra dimension $y$ \cite{ArkaniHamed:1998sj}. Also, lepton number violation (LNV) can be maximally broken at a scale $\Lambda_{LNV}$ on a LNV brane in the extra dimension. Following the scenario as described in \cite{ArkaniHamed:1998vp}, the information of this breaking is transmitted by a bulk field $\chi$ that decreases exponentially as the distance to the LNV brane rises:
\be <\chi> \propto \e^{-m r} \ee
where $m$ is the mass of the messenger field and $r$ its distance to the LNV brane.

\begin{figure}
 \centering
 \includegraphics[width=\linewidth]{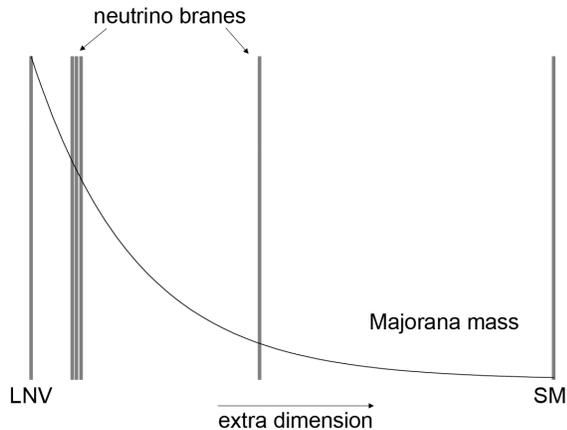}
 \caption{Evolution of the Majorana mass from the LNV brane to the SM brane along the extradimensional bulk and the four branes where the four neutrino generations are localized.}
 \label{fig:ExtraDimension}
\end{figure}

We now locate the right-handed neutrino of each generation on a different brane along the extra dimension. The overlap of the neutrino wave function and the messenger field is different for each generation such that each generation sees a different amount of LNV. This setup is sketched in Fig.~\ref{fig:ExtraDimension}.

The information of LNV is given by the Majorana mass terms of the four neutrino generations, and thus an exponential ansatz for the Majorana masses along the extra dimension is chosen:
\begin{equation}
M_{R} (y) = \Lambda_{LNV} \e^{-y} ,
\end{equation}
where $y$ is the axis along the extra dimension. 

The effective Majorana mass for the neutrino of generation $i$ is
\be
M_{Ri} = \Lambda_{LNV} \e^{-\alpha_i}.
\ee

We choose the Yukawa couplings $y_i$ equally distributed for each generation,
\be
y_4 - y_3 = y_3 - y_2 = y_2 - y_1 = 0.25
\ee
with $y_4 = 1$.

The Majorana masses are then constrained by neutrino oscillation data ($\Delta m_{12}^2$ and $\Delta m_{13}^2$) for the first three generations and by Fig.~\ref{fig:Mdiff} for the fourth generation:
\be
\delta m = M_4 - m_4 = M_{R4}.
\ee

The localizations of the neutrino branes are listed in Table~\ref{tab:ED}.

\begin{table}[h]
	\centering
		\begin{tabular}{|c|c|c|c|}
\hline
$i$	&	$\alpha_i$	&	$m_i$ [GeV]		&	$M_i$ [GeV] 		\\
\hline
4	&	$\sim 43.7$	& 246				&	247			\\
3	&	9.6		& $5.0\cdot 10^{-11}$		& $6.7\cdot 10^{14}$		\\
2	&	8.7		& $9.1\cdot 10^{-12}$		& $1.7\cdot 10^{15}$		\\
1	&	$\leq 7.9$	& $\leq 4.1\cdot 10^{-12}$	& $\geq 3.7\cdot 10^{15}$	\\
\hline
		\end{tabular}
	\caption{Localizations of the neutrino branes in the extradimensional bulk and corresponding mass eigenvalues for $\Lambda_{LNV} = \unit[10^{19}]{GeV}$.}
	\label{tab:ED}
\end{table}

The positions of the neutrino branes soften the hierarchy of the neutrino masses in a significant way. Thus, in an extradimensional framework the huge gap between the first three and the fourth generation is considerably smaller.

\section{Summary}
We have revisited bounds on additional Majorana neutrinos with the assumption of finite mixing to the electron neutrino, in order to provide a useful guide for fourth generation neutrino model building. We have shown that a fourth generation Majorana neutrino is not yet excluded if it has a mass of several hundred GeV and the Majorana states pair up to form a pseudo-Dirac state. The mixing of such a  neutrino is dominantly constrained by the radiative decay of the muon. Because of the pseudo-Dirac nature, lepton number violating processes like like-sign dilepton production turn out to be strongly suppressed. Besides being potentially observable in next generation $\pr$ experiments, the pseudo-Dirac neutrinos could be directly produced at the LHC, as discussed in \cite{delAguila:2008hw}. In this paper a 5~$\sigma$ discovery reach for heavy neutrino
masses up to 100~GeV was advocated with 30~fb$^{-1}$. While for larger masses the production cross section would decrease, new decay channels open up once 
the heavy neutrino mass exceeds the Higgs mass, which would require a detailed simulation. Finally we have shown that the large mass hierarchy can be softened within extradimensional models.

\section*{ACKNOWLEDGEMENTS}
We thank J. Herrero-Garcia, A. Aparici, N. Rius, and A. Santamaria for pointing out an error in the first draft of this manuscript \cite{Lenz:2010ha} and Paul Langacker for valuable discussions.

\end{document}